\documentclass[prl,twocolumn,showpacs,superscriptaddress,
groupeaddress, preprintnumbers,amsmath,amssymb,tightenlines]{revtex4}
\usepackage{graphicx}
\newcommand{\beq}{\begin{equation}}
\newcommand{\eeq}{\end{equation}}
\newcommand{\bea}{\begin{eqnarray}}
\newcommand{\eea}{\end{eqnarray}}
\newcommand{\ba}{\begin{array}}
\newcommand{\ea}{\end{array}}
\newcommand{\bc}{\begin{center}}
\newcommand{\ec}{\end{center}}
\newcommand{\lsimeq}{\alt}
\newcommand{\gsimeq}{\agt}

\newcommand{\bml}{\begin{subequations}}
\newcommand{\eml}{\end{subequations}}
\newcommand{\commentout}[1]{{}}
\newcommand{\bk}{{\bf k}}

\newcommand{\K}{{\cal K}}
\newcommand{\adag}{a^\dagger}
\newcommand{\alphadag}{\alpha^\dagger}
\newcommand{\ddt}{\frac{d}{dt}}
\newcommand{\half}{\hbox{$\frac{1}{2}$}}

\newcommand{\HC}{{\rm H.c.}}
\newcommand{\eq}[1]{(\ref{#1})}
\newcommand{\etal} {{\it et al.\/}}

\newcommand{\vol}[1]{{\bf #1}}
\newcommand{\comment}[1]{{}}

\begin{document}
\title{Bose-Einstein Condensate of Trimers
  Dressed by Atom-Dimer Cooper Pairs}
\author{Matt Mackie}
\affiliation{Department of Physics, University of Turku, FIN-20014 Turun
yliopisto, Finland}
\affiliation{Helsinki Institute of Physics, PL 64, FIN-00014
Helsingin yliopisto, Finland}
\author{Jyrki Piilo}
\affiliation{Department of Physics, University of Turku, FIN-20014 Turun
yliopisto, Finland}
\affiliation{Institute of Solid State Physics, Bulgarian Academy
of Sciences, Tsarigradsko chauss\'{e}e 72, 1784 Sofia, Bulgaria}
\author{Olavi Dannenberg}
\affiliation{Helsinki Institute of Physics, PL 64, FIN-00014
Helsingin yliopisto, Finland}
\date{\today}

\begin{abstract}
We theoretically examine the neutral atom-molecule analogue of the
anomalous quantum correlations between degenerate electrons, i.e., Cooper
pairs, that are responsible for superconductivity. Based on rogue
dissociation of triatomic molecules (trimers) into opposite-momentum pairs
of atoms and diatomic molecules (dimers) via a photoassociation or
Feshbach resonance, we find a superfluid transition to a Bose-Einstein
condensate of trimers dressed by atom-dimer Cooper pairs, at a critical
temperature in reach of present ultracold technology.
\end{abstract}


\maketitle

Exotic ultracold molecules are all the rage. For example, there are
prospects for quadratomic molecules via photoassociation of
ultracold polar molecules~\cite{AVD03}, and evidence of forbidden
molecular transitions in ultracold Rydberg atoms~\cite{FAR03}. Going
colder to Bose-Einstein condensates (BECs), there are opportunities for
amplified selectivity in photodissociation of triatomic
molecules~\cite{MOO02}, and mesoscopic molecular ions in BEC doped with
an charged impurity~\cite{COT02}. Lately this frenzy has
surrounded ultracold~\cite{REG03} and condensate~\cite{GRE03} diatomic
molecules emerging from a Fermi sea of atoms.

In basic magnetoassociation, a molecule is created when one atom from a
colliding pair undergoes a spin flip in the presence of a
magnetic field tuned near a Feshbach resonance. A similar molecular state
arises when two atoms absorb a photon tuned near a
photoassociation resonance. The statistics of neutral
molecules are determined by the number of
neutrons in the nuclei of the constituent atoms: odd for fermions, even
for bosons. Resonant association thus presents an unusual
opportunity to change the particle statistics, leading to cooperative
fermion behavior. For example, a photoassociation~\cite{MAC01} or
Feshbach resonance~\cite{TIM01} could deliver the neutral-atom analogue
of the anomalous quantum correlations between electrons (Cooper pairs)
responsible for superconductivity~\cite{TIN75,FRI89}.
Magnetoassociation experiments~\cite{REG03} are presently poised to
realize this regime.

Meanwhile, attention is turning to
resonant association~\cite{SCH02,DAN03} in Bose-Fermi mixtures of
atoms~\cite{TRU01}, and the subsequent atom-molecule Cooper
pairing~\cite{MAC03}. However, the most optimistic superfluid transition
temperature is an order of magnitude below what is currently
feasible~\cite{MAC03}. Based on rogue dissociation of triatomic molecules
(trimers) into opposite-momentum pairs of atoms and diatomic molecules
(dimers), we therefore report a high-temperature superfluid transition to
a Bose-Einstein condensate of trimers dressed by atom-dimer Cooper pairs.
Upon diagonalizing the Hamiltonian, we present theory on the
trimer-molecule binding energy and the superfluid transition temperature.
For $^6$Li atoms and $^7$Li-$^6$Li dimers, the superfluid transition
should occur at about a tenth the atomic Fermi temperature.

In its simplest guise, the theory of collective resonant association
is a two-mode model, analogous to second-harmonic-generated
photons, which couples zero-momentum atomic and molecular condensates.
Rogue~\cite{JAV02} transitions to the continuum of noncondensate atomic
modes~\cite{GOR01} occur because dissociation of
a zero-momentum ($\bk=0$) condensate molecule need not result in
zero-momentum condensate atoms, but may just as well deliver two
noncondensate atoms with equal and opposite momentum
($\pm\bk$). Rogue dissociation ultimately leads to anomalous quantum
correlations between atoms, which are the analogue
of Cooper pairs. An immediate consequence of said correlations is
Ramsey fringes between atoms and molecules~\cite{DON02}. The same
intuition will also prove useful here.

We model an ideal degenerate
mixture of fermionic atoms and dimer molecules coupled by either a
Feshbach or photoassociation resonance to bound trimer molecules. An
ideal gas is chosen mainly because off-resonant particle-particle
interactions are generally too weak for practical purposes. The initial
fermionic atom-dimer state could be prepared using a
Raman scheme for photoassociating a degenerate Bose-Fermi mixture of
atoms~\cite{MAC03}, and selectively removing the leftover bosons. The
atom-dimer $\leftrightarrow$ trimer resonance is expectedly well resolved
so that, once the initial atom-dimer state has been created, transitions
involving three free atoms are avoided. In contrast to the all-boson
case~\cite{MOO02}, ultracold transitions that involve a free
bosonic atom are Pauli blocked, i.e., the two identical fermionic
constituents of a given trimer may not form a bound state.

In second-quantization parley, a particle of mass $m_\sigma$ and momentum
$\hbar\bk$ is described by the creation operator $a_{\bk,\sigma}$. The
greek index corresponds mnemonically to the number of constituent atoms a
given particle contains: $3$ for bosonic trimers, 2 for fermionic dimers,
and 1 for fermionic atoms. All operators obey their (anti)commutation
relations. The microscopic Hamiltonian for such a freely-ideal system is
written:
\bea
\frac{H}{\hbar} &=&
\sum_{\bk,\sigma} \left[\left(\epsilon_{k,\sigma}-\mu_\sigma\right)
  \adag_{\bk,\sigma}a_{\bk,\sigma}\right]
\nonumber\\
&&-\frac{\K}{\sqrt{V}}
  \sum_{\bk,\bk'}\left(\adag_{\bk+\bk',3}a_{\bk,1}a_{\bk',2}+\HC\right).
\label{MICRO_HAM}
\eea
The free-particle energy is defined by
$\hbar\epsilon_{k,\sigma}=\hbar^2 k^2/2m_\sigma$, and the chemical
potential by $\hbar\mu_\sigma$. In particular, the molecular
chemical potential is defined by
$\mu_3=2\mu-\delta_0$, where the bare detuning $\delta_0$ is a measure of
the binding energy of the trimer with
$\delta_0>0$ taken as above threshold. The (mode-independent)
atom-molecule coupling is $\K$, and
$V$ is the quantization volume.

The key realization is how to cast the Hamiltonian~\eq{MICRO_HAM} into a
readily diagonalized form. Consider a {\em time-dependent} unitary
transformation, which leaves the physics unchanged providing
$
H\rightarrow U^\dagger HU-iU^\dagger\partial_tU.
$
Given the generator
$U=\Pi_{\bk,\sigma}\exp[-it u_{\bk,\sigma}\adag_{\bk,\sigma}a_{\bk,\sigma}]$,
then $u_{\bk,3}=u_{\bk,1}+u_{\bk,2}$ implies
$[H,U]=0$ and, thus, $H\rightarrow H-iU^\dagger\partial_tU$.
Appropriately armed, apply the unitary transformation specified by
$u_{\bk,1(2)}=[\epsilon_{k,1(2)}-\epsilon_{k,2(1)}]/2$, which conveniently
corresponds to the special case
$u_{\bk,3}=0$, leaving the trimer term unchanged. The new
Hamiltonian reads:
\bea
\frac{H}{\hbar} &=&
\sum_\bk \left[\left(\epsilon_{k,3}+\delta_0-2\mu\right)\adag_{\bk,3}
a_{\bk,3}
  +(\varepsilon_k-\mu)\adag_{\bk,\sigma} a_{\bk,\sigma}\right]
\nonumber\\&&
-\frac{\K}{\sqrt{V}}\sum_{\bk,\bk'}\left(\adag_{\bk+\bk',3}
a_{\bk,1}a_{\bk',2}
  +\HC\right),
\label{HAM}
\eea
where the reduced free-particle energy is
$\hbar\varepsilon_k=\hbar^2 k^2/4m^*$, with $1/m^*=1/m_1+1/m_2$. 
Also, chemical equilibrium has been incorporated as $2\mu=\mu_1+\mu_2$. We
may now make a transformation to a dressed basis:
\bml
\beq
\left(\begin{array}{c}\alpha_{\bk,1}\\ \alphadag_{-\bk,2}\end{array}\right)
  =\left(\begin{array}{cc}
    \cos\theta_k & -e^{i\varphi}\sin\theta_k \\
    e^{-i\varphi}\sin\theta_k & \cos\theta_k
  \end{array}\right)\!\!
  \left(\begin{array}{c} a_{\bk,1}\\ \adag_{-\bk,2}\end{array}\right)\!\!,
\label{BOGOa}
\eeq
\beq
\alpha_{\bk,3} = a_{\bk,3} +\sqrt{V}\Phi\delta_{\bk,0},
\label{BOGOb}
\eeq
\label{BOGOT}
\eml
where $\delta_{\bk,0}$ is the Kronecker delta-function,
\bea
\frac{H}{\hbar} &=& \left(\delta_0 -2\mu\right)V|\Phi|^2 
  +\sum_\bk
    \left(\epsilon_{k,3}+\delta_0-2\mu\right)\alphadag_{\bk,3}\alpha_{\bk,3}
\nonumber\\&&
  +\sum_\bk \left[ \left( \varepsilon_k -\mu\right) 
    +\omega_k\left(\alphadag_{\bk,1}\alpha_{\bk,1}
      +\alphadag_{\bk,2}\alpha_{\bk,2}-1\right)\right].
\nonumber\\
\label{DIAG_HAM}
\eea
The condensate mean-field is
$\langle a_{0,3}\rangle/\sqrt{V}=e^{i\varphi}|\Phi|$, the mixing angle is
$\tan2\theta_k=|\Phi|\K/(\varepsilon_k-\mu)$, the quasiparticle frequency
is $\omega_k^2=(\varepsilon_k -\mu)^2 +|\Delta|^2$, and the gap
is $|\Delta|=\K|\Phi|$. The Hamiltonian~\eq{MICRO_HAM} is now lowest-order
diagonal.

Since the contribution of rogue dissociation
to resonant association can be written as the set of all one-loop Feynman
diagrams, and since applying the canonical transformation~\eq{BOGOT} is
in fact equivalent to summing over that set~\cite{FRI89}, then our generic
intuition~\cite{JAV02} is clearly applicable. Broadly put, the
lowest-energy configuration of the system is a molecular Bose-Einstein
condensate dressed by anomalously correlated pairs of equal and opposite
momentum. More precisely, sweeping the detuning continuously evolves this
superposition from all trimers well below threshold, to an admixture of
trimers and atom-dimer Cooper pairs near threshold, to all atom-dimer
Cooper pairs well above threshold.

To illustrate, consider the mean-field Heisenberg equations
for the bosonic operator $a_{0,3}$ and the anomalous-pair-correlation
operator
$C_\bk={a}_{\bk,1}{a}_{-\bk,2}$ ({\it sans} chemical potential and
collective enhancement):
\bml
\bea
i\ddt\langle{a}_{0,3}\rangle&=&\delta_0\langle a_{0,3}\rangle
  -\frac{\K}{\sqrt{V}}\sum_{\bk} \langle C_\bk\rangle,
\label{HEIS_EQa}
\\
i\ddt\langle C_\bk\rangle
  &=&2\varepsilon_k\langle C_\bk\rangle
    -\frac{\K}{\sqrt{V}}\langle a_{0,3}\rangle.
\label{HEIS_EQb}
\eea
\label{HEIS_EQ}
\eml
Below threshold, simple Fourier analysis delivers the binding
energy, $\hbar\omega_B<0$, of the Bose-condensed trimers:
\beq
\omega_B-\delta_0-\Sigma(\omega_B)+i\eta=0,
\label{BIND}
\eeq
where $\Sigma(\omega_B)$ is the self-energy of the Bose molecules and
$\eta=0^+$. Incidentally, we show elsewhere that the real
poles of equation~\eq{BIND} fit the Regal \etal~\cite{REG03} data for
the binding energy of $^{40}\rm{K}_2$ molecules, and similar
measurements for a system of trimers would uniquely determine the
parameters of the present theory. On the other side of threshold,
the critical temperature for the transition to
effectively all superfluid atom-dimer pairs is derived from
Eq.~\eq{HEIS_EQa}:
\beq
T_c/T_F\simeq \exp\left(-\frac{\pi/4}{k_F|a_R|}\right).
\label{BCS}
\eeq
Here the resonant atom-dimer scattering length is
$a_R=-(4\pi m^*/\hbar)\K^2/\delta_0$. Also, we have taken a single Fermi
wavevector, $k_F$, for the atoms and the dimers, i.e.,
$\mu_1+\mu_2=\mu_{1(2)}m_{1(2)}/m^*$; assuming that the particles see the
same trap, adjusted for mass differences, equal Fermi wavevectors are
realized if the number of atoms and dimers satisfy
$N_2/N_1=(m_1/m_2)^{3/2}$. The effective Fermi temperature is
$k_B T_F/\hbar=(\mu_1+\mu_2)m^*/\sqrt{m_1m_2}$,
or $T_F=T_F^{(1)}\sqrt{m_1/m_2}$. Last but not least, it is easy to show
$|\Delta|\propto T_c$, so that
$|\Phi|\propto\exp\,(-\pi/4k_F|a_R|)$, and the trimer part of the
dressed BEC-pair admixture becomes larger near threshold
(increasing $|a_R|$), as expected.

Whereas the below-threshold regime of a trimer condensate is no doubt of
interest ($a_R>0$), both as a precursor to fermionic superfluidity and in
its own right, we keep our focus on attractive systems. The strongly
interacting regime is defined by
$k_F|a_R|\sim1$, indicating a transition to
predominantly atom-dimer Cooper pairs  at the critical temperature
$T_c\sim 0.45T_F$. Using
$m_3=m_1+2m_1$ as an example, which is akin to a system of
$^{6}\rm{Li}$ atoms and $^{7}\rm{Li}$-$^{6}\rm{Li}$ dimers, the required
dimer-atom fraction is $N_2/N_1=0.31$, the ratio of the effective
and atomic Fermi temperatures $T_F=0.7T_F^{(1)}$, and the critical
temperature
$T_c\sim 0.3T_F^{(1)}$. Although Eq.~\eq{BCS} is of dubious validity for
$T_c\lsimeq T_F$, it confirms that resonant association should in
principle drive superfluid pairing between atoms and dimer molecules at
transition temperatures within reach of present ultracold technology.

To rigorously identify the critical
temperature for the superfluid transition, it is necessary to go beyond
the effective atom-dimer theory, and explicitly include the bosonic
molecular state. Continuing to focus on a system of
$^{6}\rm{Li}$ atoms and $^{7}\rm{Li}$-$^{6}\rm{Li}$ dimers, we return to
the Hamiltonian~\eq{DIAG_HAM} and set $\epsilon_{k,3}\approx
\half\varepsilon_k$. We also introduce a
second molecular state, which can arise
because large detuning from one state brings the system into the
neighborhood of another bound state, or because of the presence of a
scattering resonance. The Hamiltonian~\eq{DIAG_HAM} is adapted simply by
making the substitution $\delta_0\rightarrow\delta_{0,l}$
($\Phi\rightarrow\Phi_l$, $\K\rightarrow\K_l$), and summing over the
index $l$; also, the gap becomes $|\Delta|=\K_1|\Phi_1|+\K_2|\Phi_2|$.
Here $\K_2\gsimeq\K_1$, and the system is tuned between the two levels so
that
$\delta_2>0$ and $\delta_1<0$. The thermodynamic pressure is the
centerpiece of this calculation, and is obtained from the partition
function
$\Xi=\text{Tr}\exp\left( -\beta H\right)$:
\bea
p &=&
  -\sum_l\left( \delta_{0,l} -2\mu\right)|\Phi_l|^2
    +V^{-1}\sum_\bk \left( \omega_k +\mu -\varepsilon_k\right)
\nonumber\\&&
  +2(\beta V)^{-1}\sum_\bk\ln\left( 1 +e^{-\beta \omega_k}\right)
\nonumber\\&&
  -(\beta V)^{-1}
    \sum_{\bk,l}\ln\left\{ 1 -\exp\left[\beta\left(2\mu -\delta_{0,l}
      -\half\varepsilon_k\right)\right]\right\},
\nonumber\\
\eea
where $1/\beta=k_BT/\hbar$. The chemical
potential is determined from the condition
$\left.\partial p/\partial|\Phi_l|\right|_{\mu,T}=0$:
\bml
\beq
(\delta_l-2\mu)|\Phi_l| = 2\Sigma_l(0)+
  \frac{\K_l|\Delta|}{2V}\,\sum_\bk\,
    \frac{1}{\omega_k}\tanh\half\beta\omega_k.
\label{CHEMPOT}
\eeq
Renormalization is via the resonant self-energy
$\Sigma_l(0)$, meaning the summation is
ultraviolet convergent and the physical detuning
$\delta_l$ replaces the bare detuning.
Given the chemical potential, the density of the system
$\rho\equiv\langle N\rangle/V$ determines the gap according to 
$\rho=\left.\partial p/\partial\mu\right|_{T,\Phi}$:
\bea
\rho &=& 2\sum_l|\Phi_l|^2+\frac{2}{V}\sum_{\bk,l}
  \frac{1}{\exp\left[\beta\left(\half\varepsilon_k 
      +\delta_l-2\mu\right)\right] -1}
\nonumber\\&&
  +\frac{1}{V}\sum_\bk\frac{\omega_k +\mu -\varepsilon_k 
    +\left( \omega_k -\mu
+\varepsilon_k\right)\exp(-\beta\omega_k)}
{\omega_k\left[1+\exp(-\beta\omega_k)\right]}\,.
\nonumber\\
\label{DENS}
\eea
\label{ALGEBRAIC}
\eml

\begin{figure}
\centering
\includegraphics[width=8.0cm]{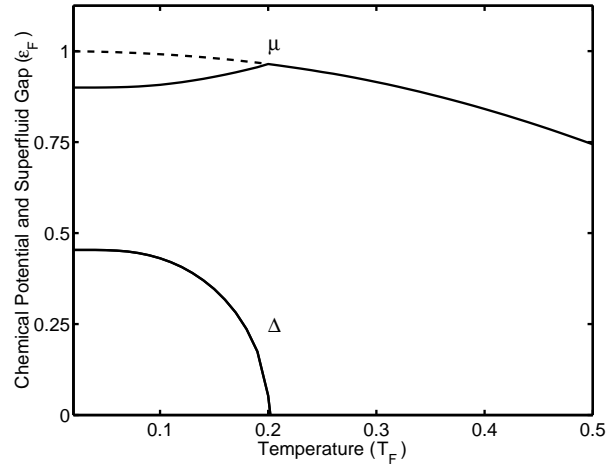}
\caption{Onset of the superfluid transition to a Bose-Einstein condensate
of trimers dressed by anomalous atom-dimer pairs. The non-zero
superfluid gap clearly lowers the system energy compared to the normal
state (dashed curve).
When the detuning is large and positive  ($\delta_2\gg\sqrt{\rho}\K_2$),
such as here, the system is mostly rogue pairs with a negligible fraction
of trimers.}
\label{SUPER}
\end{figure}

The algebraic system~\eq{ALGEBRAIC} is
sufficient to numerically determine the chemical potential as a function
of temperature, which should display a characteristic cusp at the onset
of superfluidity. Physically, a cusp arises because the superfluid
BEC-pair dressed state is lower in energy than the normal state,
implying the concurrent appearance of a non-zero gap. This intuition is
confirmed in Fig.~\ref{SUPER}. For positive detunings
large compared to the collective-enhanced coupling
($\delta_2\gg\sqrt{\rho}\K_2$ with $\delta_1\approx-\delta_2$), the
effective atom-dimer theory~\eq{BCS} with
$k_F|a_R|=1/2$ is an excellent working approximation. 
Also, the fraction of trimer is puny ($\sim 10^{-7}$), as per the large
detuning. Any $s$-wave collisional interactions are negligible
compared to the detuning, justifying the ideal-gas assumption.
The trap, albeit omitted, should actually favor the occurrence of
superfluid pairing~\cite{OHA02}.

Dimer molecules created near a Feshbach resonance are highly vibrationally
excited and, thus, characteristically long-range (K\"ohler
\etal~\cite{DON02}). Fermion-composite dimers are consequently
long-lived due to Pauli-suppressed vibrational
relaxation~\cite{PET03}, and there is no reason to expect otherwise
from Feshbach trimer states. In photoassociation, a two-color Raman
scheme is required to avoid spontaneous decay: a laser couples the atoms
to an electronically-excited intermediate trimer state, a second laser
couples the system to a ground-electronic trimer manifold, and the
intermediate trimer state is well-detuned. Long-range states are
available with photoassociation, although a Raman scheme also allows
access to stable lower-lying vibrational levels, which are much smaller
and less understood. Nevertheless, the molecular
fraction is negligibly small when the system is well above the
appropriate threshold, diminishing the chance for spontaneous
for decay of any kind.

For a signature, note that Cooper pairing was suggested to explain how
nuclei with even numbers of nucleons can have a larger excitation energy
than nuclei with odd numbers of nucleons~\cite{BOH58}.  The presence of
anomalous atom-dimer correlations should similarly blueshift the resonant
frequency of a dimer-dissociating photon by an amount proportional to the
gap, in contrast to the otherwise~\cite{PRO03} red-shifted
photodissociation spectrum (see also boson blueshift~\cite{MAC03b}).
Physically, the increased frequency comes about because, in order to
dissociate a dimer molecule, an extra amount of energy, specifically, the
gap energy, is needed to break any anomalous correlations. If the
dissociating light is weak enough to justify lowest-order free-bound
transitions, the blueshift should dominate the redshift.

In conclusion, we reinforce the idea that statistics need not be an issue
in resonant association: each of the systems (Bose, Fermi, Bose-Fermi) is
described by the same basic Heisenberg equations of motion, and will
respond as a unit to form molecules in mostly-complete cooperation.
Moreover, while it is not entirely clear how many constituents the
association bound states may in practice contain, the limit is certainly
not two. Granted, at this stage it is difficult to predict timescales,
but in this respect the quantum optics approach--based on ratios of
parameters not parameters--has a decided advantage. When the detuning is
large compared to the collectively enhanced coupling, the superfluid
transition to a BEC of trimers dressed by atom-dimer Cooper pairs can
occur at about a tenth of the atomic Fermi temperature. A molecular
condensate of trimers could be the next milestone, whereas
high-temperature anomalous pairing between different chemical species
opens the door to further analogies with condensed-matter and nuclear
physics.

In preparation of this manuscript, magnetoassociation-induced atomic
Cooper pairing was observed~\cite{REG04}.

The authors thank Juha Javanainen and Kalle-Antti Suominen for
discussions, and the
Academy of Finland (JP and MM, project 50314), the EU network COCOMO (JP,
contract HPRN-CT-1999-00129), and the Magnus Ehrnrooth Foundation (OD) for
financial support.

\end{document}